\documentclass[prb,preprint]{revtex4}
\usepackage{amsmath}

\usepackage{graphicx}
\usepackage{enumerate}

\begin{document}

\title{Modeling protein synthesis from a physicist's perspective: A toy model} 
\author{Aakash Basu}
\affiliation{Department of Physics, Indian Institute of Technology,
Kanpur 208016, India}

\author{Debashish Chowdhury}
\email{debch@iitk.ac.in}
\affiliation{Department of Physics, Indian Institute of Technology,
Kanpur 208016, India}

%\date{\today}

\begin{abstract}
Proteins are polymers of amino acids. These macromolecules are 
synthesized by intracellular machines called ribosomes. Although 
the experimental investigation of protein synthesis 
has been a traditional area of research in molecular cell biology, 
important quantitative models of protein synthesis have been reported 
in research journals devoted to statistical physics and 
related interdisciplinary topics. From the perspective of a physicist, 
protein synthesis is the classical transport of 
interacting ribosomes on a messenger RNA (mRNA) template that dictates 
the sequence of the amino acids on the protein. We discuss
appropriate simplification of the models and methods. In particular, 
we develop and analyze a simple toy model using some elementary 
techniques of non-equilibrium statistical mechanics and predict the 
average rate of protein synthesis and the spatial organization of 
the ribosomes in the steady state. 
\end{abstract}


\maketitle

\section{Introduction}

Physical frontiers in biology\cite{ajpresource} and biological frontiers
of physics\cite{phystoday} are now active areas of interdisciplinary 
research. There are journals such as {\it Physical Biology} whose aim is to 
foster ``the integration of biology with the traditionally more 
quantitative fields of physics,...'' \cite{physbiol}. Biological physics is also one of 
the interdisciplinary topics on which papers are 
published regularly in high-impact research journals such as {\it Physical 
Review Letters}. However, often the work is too 
technical to be accessible to those who are not expert in 
modeling of biological systems. 

The main aim of this paper is to bring a piece of 
contemporary research into the classroom by appropriate simplification 
of the models and methods. In particular, we develop a simple model 
for the collective movement of ribosomes when these macromolecular 
machines move along a template messenger RNA (mRNA) strand, each 
separately synthesizing one copy of the same protein. 
\cite{spirinbook,frank06} In spite of its simplicity, this toy 
model captures the most essential steps in the process of protein 
synthesis. Because of the simplicity of the model and the pedagogical 
presentation of the calculations, even senior undergraduate students 
can obtain a glimpse of a frontier of current interdisciplinary research 
involving biology and physics.

Because of the interdisciplinary
nature of the topic, we present in Sec.~\ref{bio} a summary of the essential biochemical and mechanical
processes involved in protein synthesis. In Sec.~\ref{models} we 
present the model and highlight its main features. We report our 
results in Secs.~\ref{extended} and \ref{openbc}. We compare the 
model and the results with those for some other similar systems 
in Sec.~\ref{comparison}. Finally, in Sec.~\ref{conclusion} we 
draw our conclusions.

\section{\label{bio}Protein synthesis: essential mechano-chemical processes}

A protein is a linear bio-polymer whose monomeric subunits, called 
amino acids, are linked together by peptide bonds. A polypeptide, a 
precursor of a protein, is synthesized from the corresponding messenger 
RNA (mRNA) template by a machine called ribosome.\cite{spirinbook,frank06}
An mRNA is also a linear bio-polymer whose monomeric subunits are the 
nucleotides. Triplets of nucleotides form one single codon. The sequence 
of amino acids in a polypeptide is dictated by the sequence of codons in 
the corresponding mRNA template. 

\begin{figure}[h!] 
\begin{center}
\includegraphics[width=0.9\columnwidth]{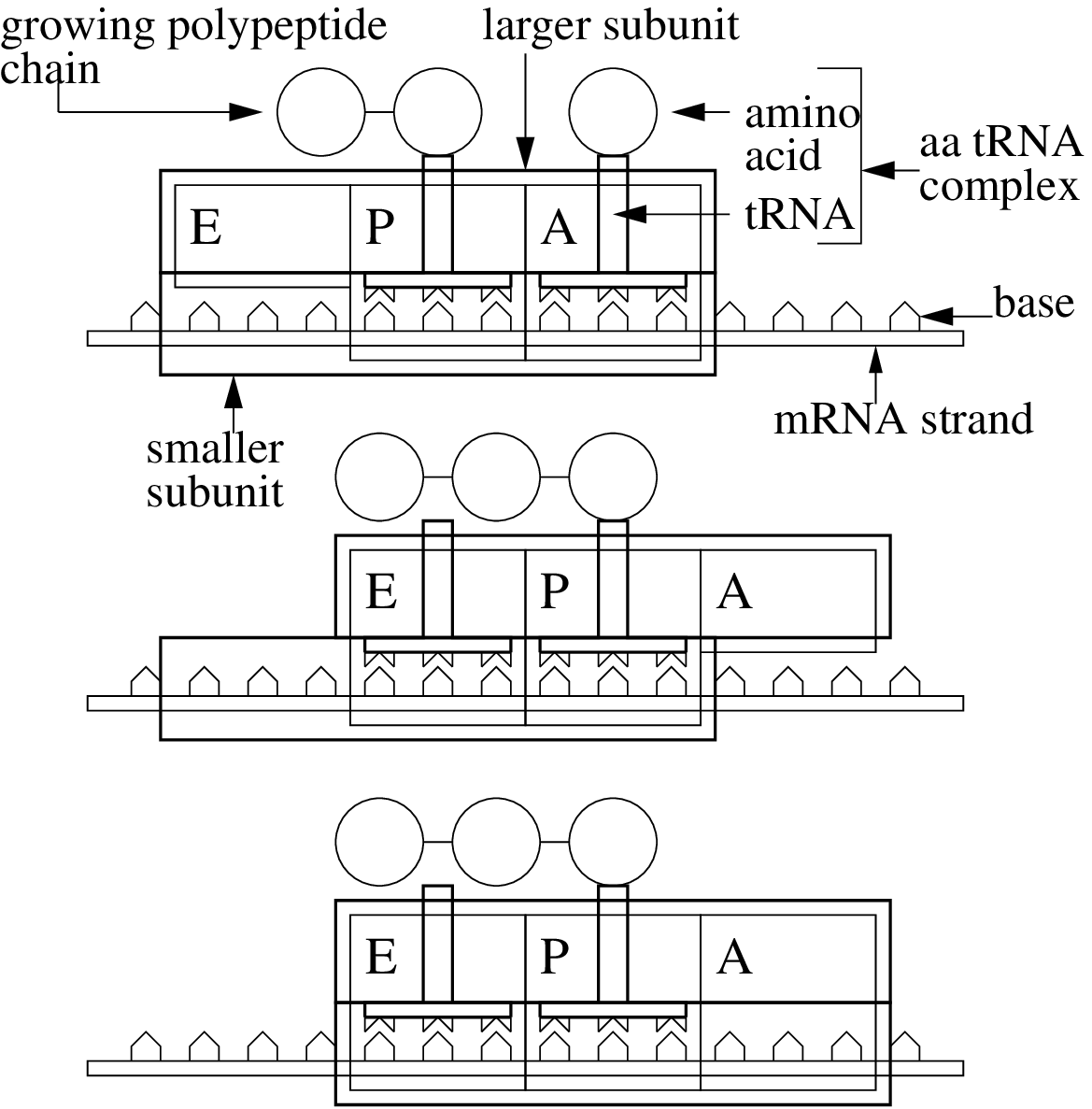}
\end{center}
\caption{Cartoons showing the three major steps in the mechano-chemical 
cycle of a single ribosome. The larger and smaller subunits are 
represented by two rectangles. The vertical ``tips'' and ``dips'' 
emphasize the codon-anticodon complementarity. The uppermost cartoon 
depicts a freshly selected aa-tRNA whose anticodon is complementary to 
the codon on the mRNA. The middle cartoon captures the situation where, 
following the formation of the peptide bond between the existing 
polypeptide and the newly recruited amino acid and the subsequent forward 
movement of the tRNA molecules from P and A to E and P sites respectively, 
the larger subunit has stepped ahead by one codon. The lowermost cartoon 
depicts the penultimate step of a cycle when the smaller subunit has also 
translocated to the next codon and the tRNA bound to the E site is 
about to exit the ribosome.} 
\label{fig-3steps}
\end{figure}

The amino acid, corresponding to a given codon, is delivered by an 
adapter molecule called transfer RNA (tRNA) (see Fig.~\ref{fig-3steps}). 
One end of a tRNA molecule consists of an anticodon (a triplet of 
nucleotides), and the other end carries the cognate amino acid (that is, 
the amino acid that corresponds to its anticodon). Because of the 
codon-anticodon complementarity, each codon on the mRNA becomes converted 
into a particular species of amino acid on the polypeptide. 
A tRNA molecule bound to its cognate amino acid is called 
aminoacyl-tRNA (aa-tRNA).

Each ribosome consists of two subunits. The mechano-chemical processes 
in these two subunits are coupled and maintain proper coordination for 
the overall operation of the ribosome. Each of the three binding sites 
(E, P, and A), which are located in the larger subunit of a ribosome, 
can bind to a tRNA (see Fig.~\ref{fig-3steps}). The binding site on the 
smaller subunit of the ribosome can bind to the mRNA template strand. 

Three major steps in the biochemical cycle of a ribosome are sketched 
in Fig.~\ref{fig-3steps}. In the first, the ribosome selects an aa-tRNA 
whose anticodon is exactly complementary to the codon on the mRNA. Next, 
it catalyzes the formation of the peptide bond between the existing 
polypeptide and the newly recruited amino acid resulting in the elongation 
of the polypeptide by one monomer. Finally, it completes the 
mechano-chemical cycle by translocating itself completely to the next 
codon and is ready to begin the next cycle. In the next section we
develop a toy model to capture these three steps in the chemo-mechanical 
cycle of a ribosome (see Fig.~\ref{states}).

\section{\label{models}Protein synthesis: a toy model}

In all the theoretical models,\cite{macdonald68,macdonald69,lakatos03,shaw03,shaw04a,shaw04b,chou03,chou04,dong,basuchowpre}
including our toy model proposed here, the sequence of codons on a given 
mRNA is represented by the corresponding sequence of the equispaced sites 
of a regular one-dimensional array or lattice. In all these 
models, the steric interactions among the ribosomes are taken into account 
by imposing the condition of mutual exclusion; that is, no codon can be 
covered simultaneously by more than one ribosome. 

In their pioneering work, MacDonald, Gibbs, and coworkers 
\cite{macdonald68,macdonald69} modeled each ribosome by an extended 
particle (effectively, a hard rod) of length $\ell$ in the 
units of a codon ($\ell$ is an integer). In reality, 
a ribosome is a complex macromolecular aggregate of proteins and RNA. 
It is not an inert rod, but a machine whose mechanical movements 
along an mRNA strand is coupled to its biochemical cycle.\cite{spirinbook}

Recently, we have reported\cite{basuchowpre} a detailed 
quantitative theory of protein synthesis. Our theoretical treatment is 
based on standard methods of non-equilibrium statistical mechanics.
Our model differs from earlier models in the way we capture the 
structure, biochemical cycle, and translocation of each ribosome. 
The toy model we propose here is a simplified version of the 
model developed in Ref.~\onlinecite{basuchowpre}.

\begin{figure}[h!]
\begin{center}
\includegraphics[width=0.9\columnwidth]{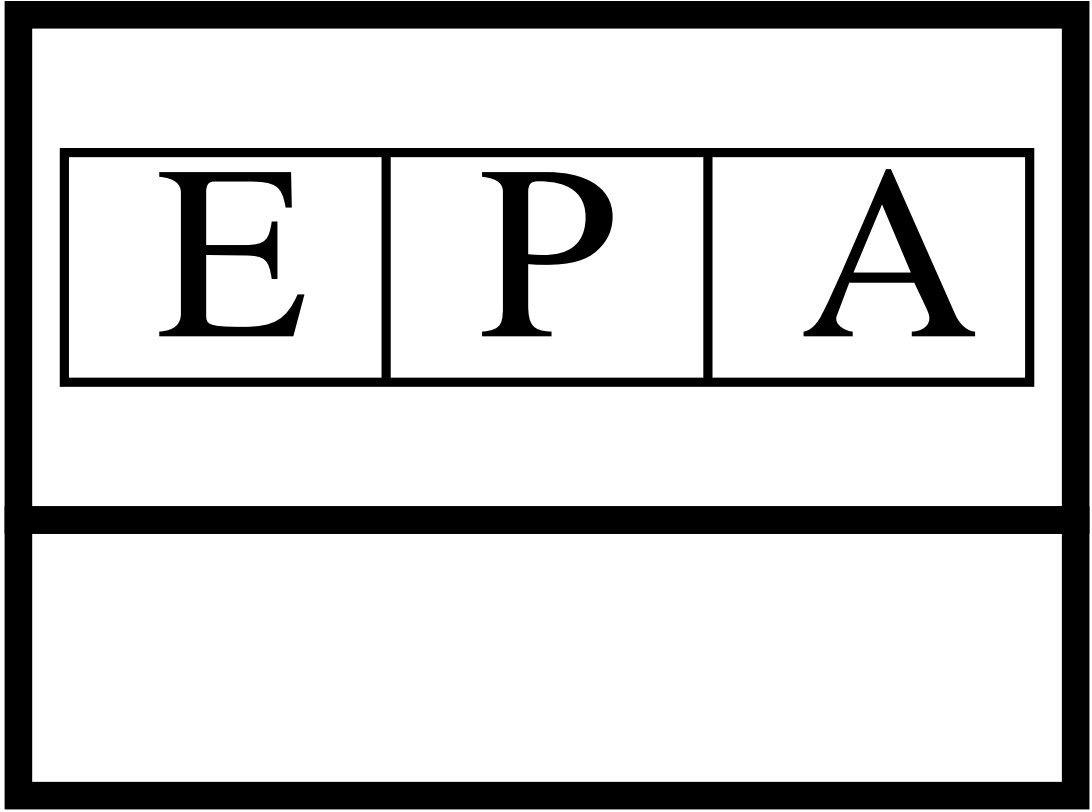}
\includegraphics[width=0.9\columnwidth]{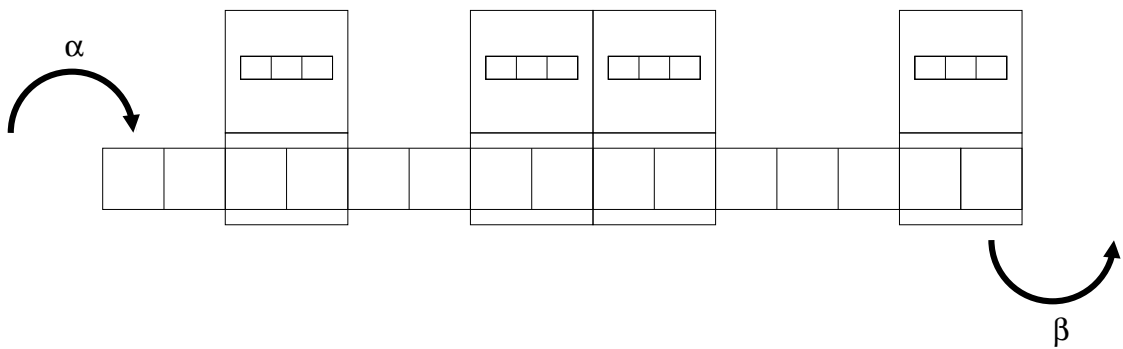}
\end{center}
\caption{A schematic representation of the model. (a) A cartoon 
of a single ribosome that explicitly shows the three binding 
sites E, P, and A on the larger subunit which is represented by 
the upper rectangle. The rectangular lower part represents the 
smaller subunit of the ribosome. (b) The mRNA is represented by 
a one-dimensional lattice where each site corresponds to a 
single codon. The smaller subunit of each ribosome covers 
$\ell$ codons ($\ell = 2$ in this figure) at a time.}
\label{fig-model}
\end{figure}

Our model is shown schematically in Fig.~\ref{fig-model}. 
We represent the single-stranded mRNA template chain, by a one-dimensional 
lattice. We label the sites of the lattice by the integer index $i$ (by 
convention, from left to right). Each of the sites from $i=1$ to $i=L$ 
represent a single codon, where $i = 1$ represents the start codon and 
$i = L$ corresponds to the stop codon. In our model, the small subunit of 
each ribosome covers $\ell$ codons at a time; the position of each 
ribosome is denoted by the integer index of the lattice site covered by 
the leftmost site of the smaller subunit. Thus, the allowed range of 
the positions $j$ of each ribosome is $1 \leq j \leq L$. No lattice site 
is allowed to be covered simultaneously by more than one overlapping 
ribosome. Irrespective of the length $\ell$, each ribosome can move 
forward by only one site in each step because it must translate successive 
codons one by one.

There are close similarities between the collective movements of the 
ribosomes along the template mRNA strand and those of vehicles on 
highways. Therefore, from the perspective of statistical physics, 
protein synthesis is also a problem of ribosomal traffic.\cite{polrev}
In the particle-hopping models of vehicular traffic,\cite{css,cnss} 
each vehicle is modeled by a particle. Moreover, a single lane of a 
highway is represented by a lattice of equispaced points (or, 
equivalently, a lattice of boxes each centered around a lattice 
site) none of which can accommodate more than one particle at a time. 
Each of these self-propelled particles can move forward by a 
maximum of $v_{\max}$ lattice sites, unless hindered by another 
vehicle in front of it.

We will compare and contrast some of the characteristic features of 
ribosomal traffic with the corresponding features of vehicular traffic. 
In analogy with vehicular traffic, we define the flux $J$ as the 
average number of the ribosomes crossing a specific codon (selected 
arbitrarily) per unit time. We borrow the terminology of traffic 
science\cite{css} and refer to the flux-density relation as 
the {\it fundamental diagram}. 

In the context of ribosomal traffic, the position, average speed, and 
flux of ribosomes have interesting interpretations in terms of protein 
synthesis. The position of a ribosome on the mRNA also gives the length 
of the nascent polypeptide it has already synthesized. The average 
speed of a ribosome is also a measure of the average rate of elongation 
of a single polypeptide. The flux of the ribosomes gives the total 
rate of polypeptide synthesis from the mRNA strand, that is, the number of 
polypeptides synthesized completely per unit time interval. 

In a real mRNA the nucleotide sequence is, in general, inhomogeneous, 
but far from random. Different codons appear on an mRNA with different 
frequencies. Moreover, in a given cell, not all the tRNA species, 
which correspond to different codon species, are equally abundant. 
It is possible to extend our toy model to capture these inhomogeneities 
following the numerical approach which we used in Ref.~\onlinecite{basuchowpre}. 
For simplicity, we will consider here only a homogeneous 
lattice. 

To test the accuracy of our approximate analytical results, 
we have also carried out computer simulations of our model. Because we 
found very little difference in the results for systems of size $L = 300$ 
and those for larger systems, all of our production runs were done for $L = 300$. We used random sequential updating. In this 
scheme, a lattice site is picked at random, and if it is occupied by the 
left edge of a ribosome the corresponding ribosome is considered for 
updating; completion of updating the states of $L$ lattice sites increases 
time by one step. This scheme of updating corresponds to the master equations 
formulated for the analytical description in our model. Each run begins 
with a random initial state, but the data for the first $5 \times 10^6$ 
time steps were discarded to ensure that the system had
reached a steady state. In the steady state, data were collected over 
the next $5 \times 10^6$ time steps. An outline of the main steps of the algorithm used 
for the simulation of the toy model for periodic boundary conditions 
is given in the Appendix. 

\section{\label{extended}Results for periodic boundary conditions}

Typically, a single ribosome itself covers about twelve codons (that is, 
$\ell = 12$), and interacts with others by mutual exclusion. 
The position of such a ribosome will be referred to by the 
integer index of the lattice site covered by the leftmost site of the 
smaller subunit. 

\subsection{Theoretical formulation under periodic boundary conditions}

Let $P_{\mu}(i)$ be the probability of finding a ribosome at site $i$, 
in the chemical state $\mu$ where $\mu = 1,2,3$ represents the three 
chemical states in each mechano-chemical cycle of the ribosome which 
is shown in Fig.~\ref{fig-3steps}. 
Hence, $P(i) = \sum_{\mu=1}^{3} P_{\mu}(i)$, 
is the probability of finding a ribosome at site $i$, irrespective of 
its chemical state. Let $P({i}|j)$  be the conditional 
probability that, given a ribosome at site $i$, there is another 
ribosome at site $j$. Then, $Q(i|j) = 1 - P(i|j)$ 
is the conditional probability that, given a ribosome at site $i$, site 
$j$ is empty. The periodic boundary conditions are somewhat 
artificial as, effectively, the mRNA takes the shape of a closed ring.

\begin{figure}[h!] 
\begin{center}
\includegraphics[width=0.9\columnwidth]{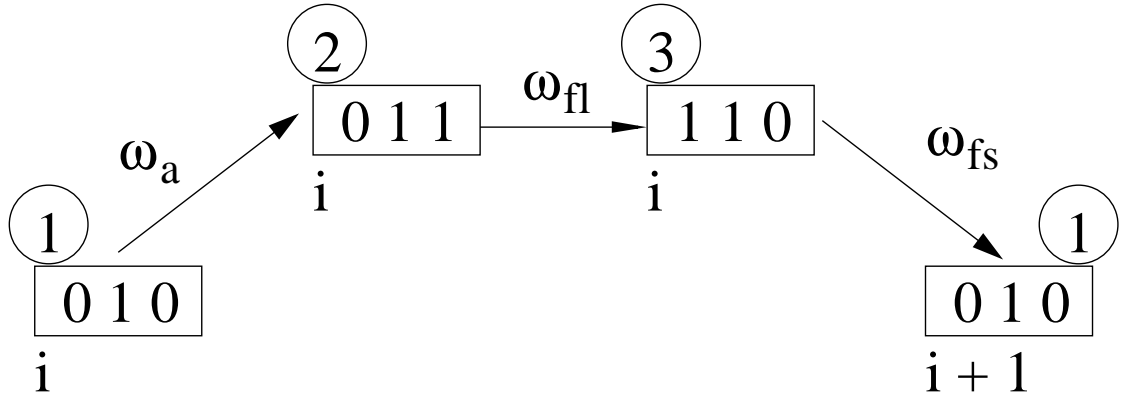}
\end{center}
\caption{A schematic representation of the simplified biochemical 
cycle of a single ribosome during protein synthesis in our toy model. 
Each box represents a distinct state of the ribosome. The integer 
index $i$ below the box labels the codon on the mRNA with which 
the smaller subunit of the ribosome binds. The number above the box 
labels the biochemical state of the ribosome. Within each box, 
$1 (0)$ represents presence (absence) of tRNA on binding sites 
E, P, A, respectively. The symbols accompanied by the arrows define 
the rate constants for the transitions from one biochemical state to 
another; $\omega_{a}$ corresponds to the selection of the aa-tRNA, and $\omega_{f \ell}$ and $\omega_{fs}$ correspond, respectively, to 
the forward movements of the large and small subunits of the ribosome. 
In our numerical calculations, we use the values of the rate constants 
for {\it E-coli} available in the literature.\cite{thompson,harrington} 
} 
\label{states}
\end{figure}

We assume that the probability of finding a ribosome at site $i$ is 
statistically independent of that of the presence or absence of other 
ribosomes at other sites. Under this approximation, the biochemical 
cycle shown in Fig.~\ref{states} implies that the corresponding 
equations for the probabilities $P_{\mu}(i)$ are 
\begin{align} \label {3:0}
\frac{\partial{}P_{1}(i)}{\partial{}t} & = \omega_{fs} P_{3}(i-1) Q(i-1|i-1+\ell) - \omega_{a} P_{1}(i), \\
\label {3:3}
\frac{\partial{}P_{2}(i)}{\partial{}t} &= \omega_{a} P_{1}(i) - \omega_{fl} P_{2}(i), \\
\label {3:4}
\frac{\partial{}P_{3}(i)}{\partial{}t} &= \omega_{fl} P_{2}(i) - \omega_{fs} P_{3}(i) Q(i|i+ \ell), 
\end{align}
respectively, where the symbols $\omega_{a}$, $\omega_{fl}$, and $\omega_{fs}$ 
are the rate constants shown in Fig.~\ref{states}. Equations of the type (\ref{3:0})--(\ref{3:4}), 
which govern the time-evolution of probabilities, are known as master equations.\cite{vankampen} The positive and negative terms on the right hand 
sides of these equations are often referred to as the gain and 
loss terms, respectively. 

The three equations (\ref{3:0})--(\ref{3:4}) are not all
independent of each other because of the condition
\begin{equation}\label {gg}
P(i)=\sum_{\mu=1}^{3}P_{\mu}(i)=\frac{N}{L} = \rho,
\end{equation}
where $\rho$ is the number density of the ribosomes on the mRNA strand. 
In our calculations, we have used 
Eqs.~(\ref{3:3})--(\ref{gg}) as the three independent equations.

For simplicity, we report here the results for only 
$\ell = 1$; the derivation of the corresponding results 
\cite{basuchowpre} for an arbitrary $\ell$ is left as an exercise 
for the reader (see Problem 1).

\subsection{Steady state properties under periodic boundary conditions}

In the steady state, all the $P_{\mu}(i)$ become independent of time. 
Because of the periodic boundary conditions, no site has any special status and the 
index $i$ can be dropped. The corresponding flux of the ribosomes $J$ 
can then be obtained from
\begin{equation}
J = \omega_{fs}P_{3}Q(i|i+ \ell),
\label{eq-fluxl}
\end{equation}
using the steady-state expressions for $Q(i|i+ \ell)$ 
and $P_{3}$.

In the special case $\ell = 1$, $Q(i|i+ \ell)$ takes the 
simple form
\begin{eqnarray} \label{3:2}
Q(i|i+1)= 1 - \rho.
\label{eq-qpbc}
\end{eqnarray}
The solution of Eqs.~(\ref{3:3})--(\ref{gg}) in the steady state for 
periodic boundary conditions is
\begin{equation}
P_{3} = \frac{\rho}{1+\Omega_{fs}(1 - \rho)},
\label{eq-p5}
\end{equation}
where,
\begin{equation}
\Omega_{fs} = \omega_{fs}/k_{\rm eff},
\end{equation}
with
\begin{equation}
\frac{1}{k_{\rm eff}} = \frac{1}{\omega_{fl}}+\frac{1}{\omega_{a}}.
\end{equation}
Note that $k_{\rm eff}^{-1}$ is an effective time that incorporates the
delays induced by the intermediate biochemical steps in between two
successive hoppings of the ribosome from one codon to the next. 
Therefore, $k_{\rm eff} \rightarrow \infty$ implies short-circuiting 
the entire biochemical pathway so that a newly arrived ribosome at 
a given site is instantaneously ready for hopping onto the next site 
with the effective rate constant $\omega_{fs}$.

If we use Eqs.~(\ref{3:2}) and (\ref{eq-p5}) in Eq.~(\ref{eq-fluxl}) 
and the definition $\rho = N/L$ for the number density, we obtain 
\begin{equation} \label{3:7}
J = \frac{\omega_{fs} \rho (1- \rho)}{1+ \Omega_{fs}(1-\rho)}.
\end{equation}
Note that $J$ vanishes at $\rho = 0$ and at $\rho = 1$ because at 
$\rho = 1$ the entire mRNA in fully covered by ribosomes. 

\begin{figure}[h!]
\begin{center}
\includegraphics[width=0.9\columnwidth]{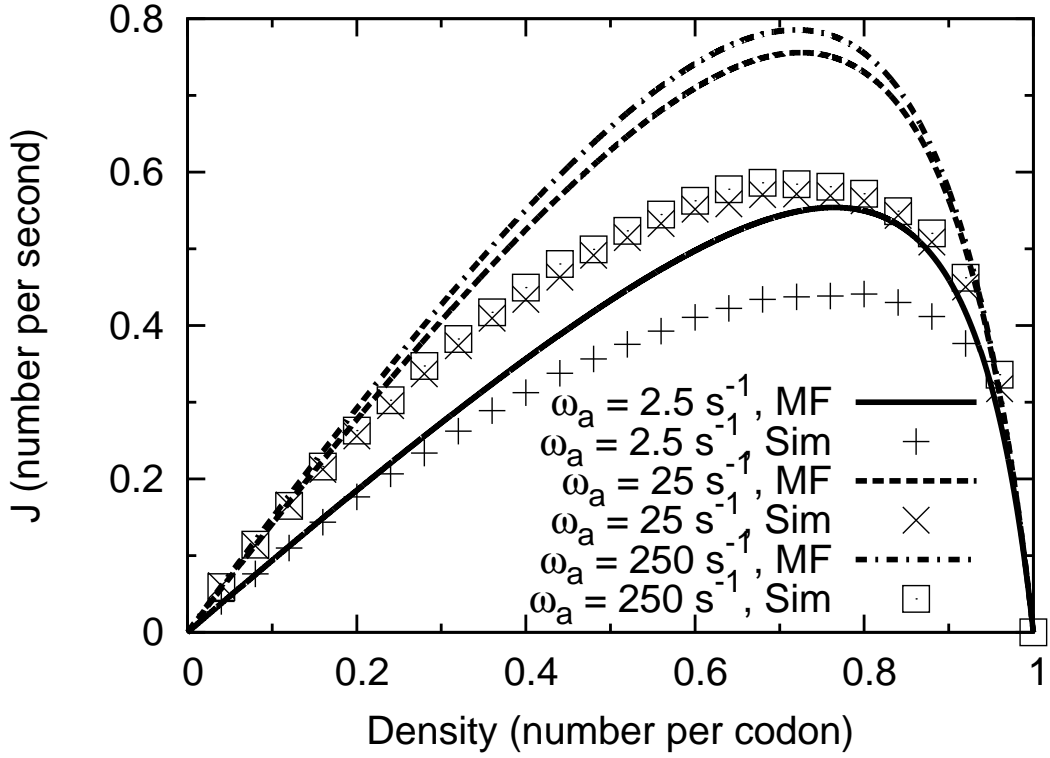}
\end{center}
\caption{Flux of ribosomes for periodic boundary conditions plotted 
against the density for three values of $\omega_{a}$. The curves 
correspond to the approximate analytical expression (\ref{3:7}), 
whereas the discrete data points were obtained by carrying out 
computer simulations. The values of all the parameters are the same as those in Table~I.} 
\label{fig-fdpbc}
\end{figure}

The flux obtained from Eq.~(\ref{3:7}) is plotted against density 
in Fig.~\ref{fig-fdpbc} for 
$\omega_{a} = 2.5$\,s$^{-1}$, $\omega_{a} = 25$\,s$^{-1}$, and 
$\omega_{a} = 250$\,s$^{-1}$. 
Comparisons of these curves with the 
corresponding simulation data (represented by discrete points in 
Fig.~\ref{fig-fdpbc}) shows that our approximate theory 
overestimates the flux. This quantitative difference, in spite of 
qualitative similarities, between our theoretical predictions and 
the simulation data arises from the correlations in the states of 
the interacting ribosomes that are neglected in our approximate 
analysis. 

The qualitative shape of the fundamental diagrams shown in 
Fig.~\ref{fig-fdpbc} for ribosomal traffic is very similar to those 
derived from similar particle-hopping models of vehicular traffic 
as well as those observed in real traffic on highways.\cite{css} 
The average flux is the product of the density and average velocity of the 
ribosomes. At very low densities, the ribosomes are sufficiently 
far apart so that each one can move freely without hindrance. In 
this regime, the average velocity remains practically unaffected by 
the increase of density and the flux increases almost linearly with 
$\rho$. As the density is increased further, the average velocity 
begins to decrease. Beyond a certain density, the average velocity 
decreases so sharply with increasing density that the overall flux 
decreases with increasing density beyond $\rho_m$, where the flux 
exhibits a maximum. We leave it as an exercise for the reader to 
extract the average velocity from the flux plotted in Fig.~\ref{fig-fdpbc} 
and to see the variation of the average velocity with $\rho$.

We next interpret the $\omega_{a}$-dependence of the fundamental 
diagrams. When $\omega_{a}$ is sufficiently small, the availability 
of the cognate tRNA molecules is the rate-limiting process; that is, 
the overall rate of protein synthesis is dominantly controlled by 
$\omega_{a}$. In contrast, when $\omega_{a}$ is so large that 
the availability of tRNA is no longer the rate limiting process, the 
flux becomes practically independent of $\omega_{a}$. Therefore, for 
a given density $\rho$, the flux increases with increasing $\omega_{a}$, 
but the rate of this increase slows down with increasing $\omega_{a}$ 
and eventually the flux saturates. 

Another interesting feature of the fundamental diagrams is the variation 
of the peak position $\rho_m$ with $\omega_{a}$. As 
the rate $\omega_{a}$ decreases, the magnitude of $\rho_m$ increases. Let us 
define $v_{\max}$ to be the maximum possible velocity of an isolated 
ribosome moving along a mRNA template unhindered by any other ribosome.
If $\omega_{a}$ is small, a ribosome has to wait on each codon for a 
longer time and the corresponding $v_{\max}$ would be low. Thus, the 
decrease of $\rho_m$ with an increase of $\omega_{a}$ can also be viewed 
as a decrease of $\rho_m$ with an increase of the effective value of $v_{\max}$ of the 
ribosomes. A similar trend for the variation of $\rho_m$ with $v_{\max}$ also has 
been observed in the fundamental diagrams of the particle-hopping
models of vehicular traffic.\cite{css} This trend is a consequence of the 
increase of the effective range of sensing mutual hindrance with 
increasing $v_{\max}$. 

\section{\label{openbc}Results for open boundary conditions}

Open boundary conditions are more realistic than periodic boundary conditions for 
modeling protein synthesis because open boundary conditions properly capture the initiation 
and termination of synthesis of proteins by each ribosome. Whenever 
the first $\ell$ sites on the mRNA in our model are vacant, this 
group of sites is allowed to be covered by a fresh ribosome with 
the probability $\alpha$ in the time interval $\Delta t$ (in all 
our numerical calculations we take $\Delta t = 0.001$\,s). Thus the effects of all the biochemical processes involved 
in the initiation of translation are captured in 
our toy model by a single parameter 
$\alpha$. Similarly, the termination of translation is also captured by 
a single parameter $\beta$; whenever the rightmost $\ell$ sites 
of the mRNA lattice are covered by a ribosome, that is, the ribosome is 
bound to the stop codon, the ribosome is detached from the mRNA 
with probability $\beta$ in the time interval $\Delta t$.
Because $\alpha$ is the probability of attachment in time $\Delta t$, 
the probability of attachment per unit time $\omega_{\alpha}$ is the solution of the equation 
$\alpha{}=1-e^{-\omega_{\alpha}\Delta t}$.
Similarly, we also define $\omega_{\beta}$ ads the probability 
of detachment of a ribosome from the stop codon per unit time.

\subsection{Steady state properties with open boundary conditions}

It is possible to do an analysis of the model with 
open boundary conditions even for arbitrary $\ell$. The method is 
similar to the one presented previously for the same model with periodic 
boundary conditions. We leave these analytical calculations 
as an exercise for the reader (see Problem~2) and present here only the results of 
computer simulations for the special case $\ell = 1$.

\begin{figure} 
\begin{center}
\includegraphics[width=0.9\columnwidth]{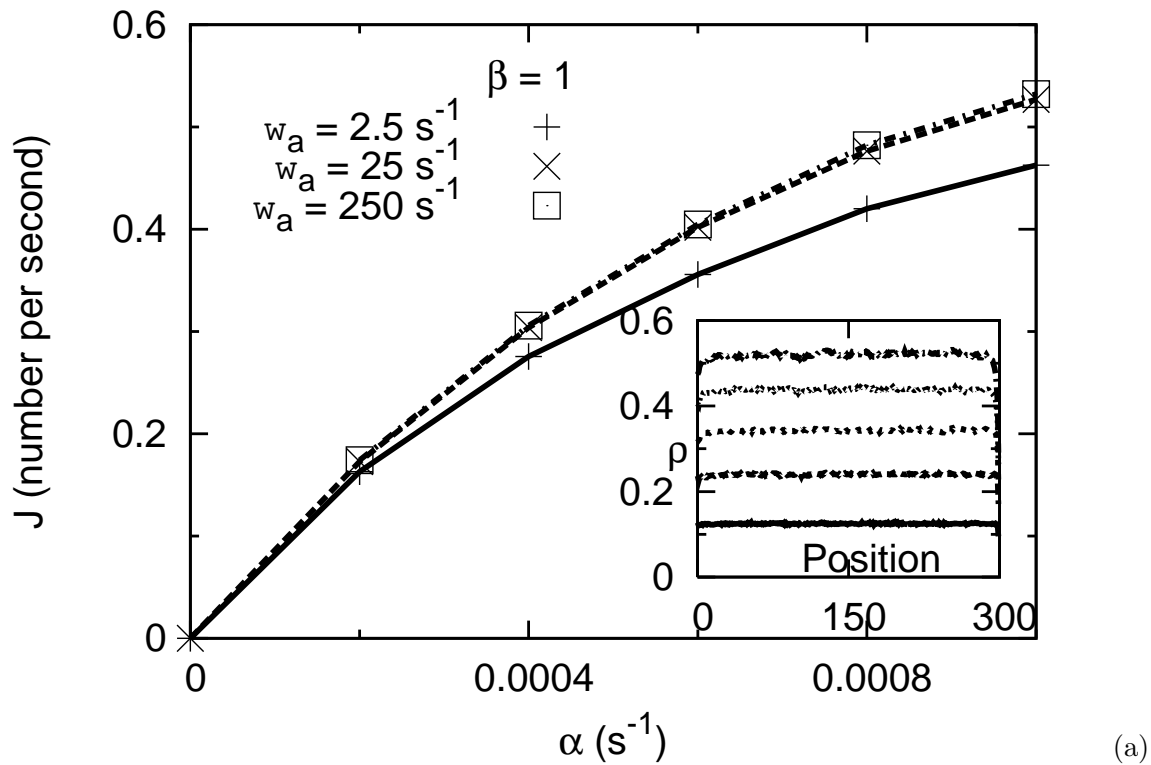}
(a) 
\includegraphics[width=0.9\columnwidth]{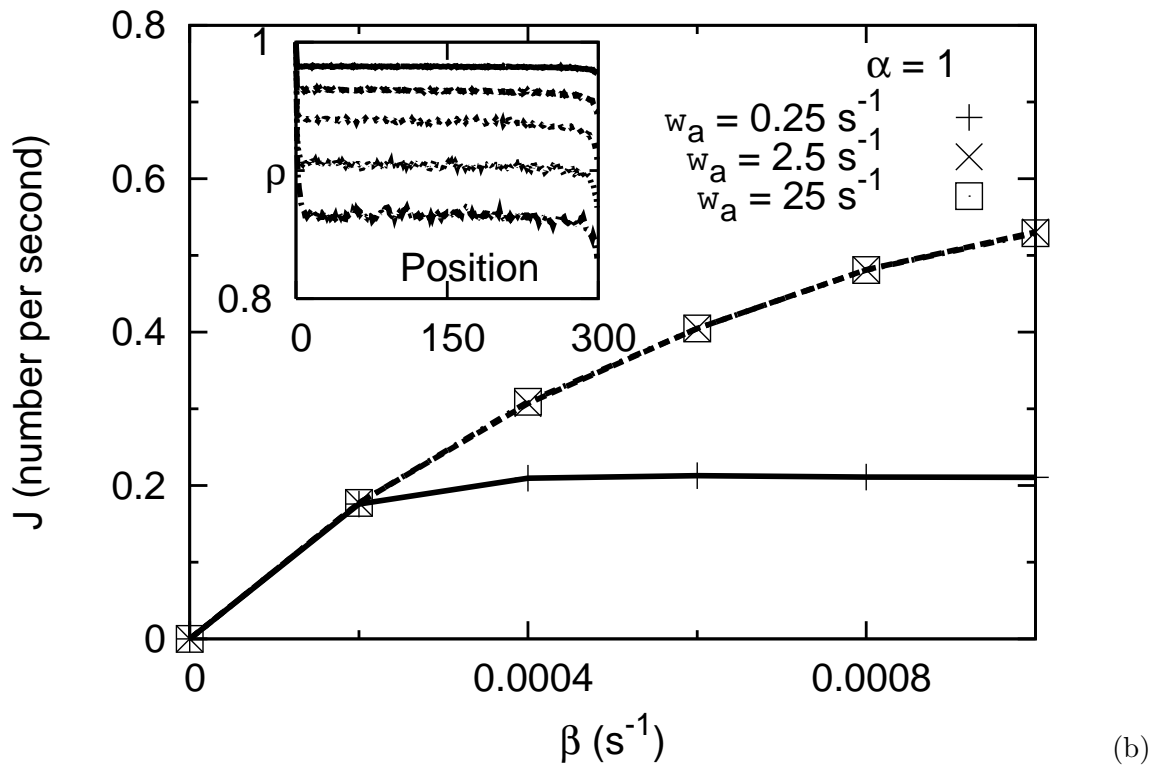}
(b) 
\end{center}
\caption{Flux of ribosomes under open boundary conditions plotted 
against $\alpha$ in (a) and $\beta$ in (b) for three 
values of $\omega_{a}$. The discrete data points were obtained 
by doing computer simulations, and the curves are merely 
guides to the eye. The average density profiles are plotted in the 
insets. In the inset of (a) the lowermost density profile corresponds 
to $\alpha=0.0002$, and the topmost one corresponds to $\alpha=0.001$; 
$\alpha$ varies from one profile to the next in steps of $0.0002$. 
In the inset of (b) the topmost density profile corresponds to
$\beta=0.0002$, and the lowermost one corresponds to $\beta=0.001$; 
$\beta$ varies from one profile to the next in steps of $0.0002$.} 
\label{fig-fluxobc}
\end{figure}

The flux $J$ found by computer simulations is plotted 
against $\alpha$ and $\beta$ in Figs.~\ref{fig-fluxobc}(a) and 
\ref{fig-fluxobc}(b), respectively. The average density profiles 
observed for several values of $\alpha$ and $\beta$ are also shown in 
the insets of Figs.~\ref{fig-fluxobc}(a) and (b). Note 
that small $\beta$ effectively creates a bottleneck at the stop 
codon and would lead to a high average density profile. In contrast, 
the ribosomes do not pile up if $\beta$ is sufficiently large. For 
$\alpha < \beta = 1$, the flux gradually increases and 
saturates as $\alpha$ increases (see Fig.~\ref{fig-fluxobc}(a)),
because a larger number of ribosomes initiate translation per unit 
time interval at higher values of $\alpha$. This increase 
of flux with increasing $\alpha$ is also consistent with the 
corresponding higher average density profile shown in the inset of 
Fig.~\ref{fig-fluxobc}(a). For $\beta < \alpha = 1$, the flux increases 
and eventually saturates with increasing $\beta$ because of the 
softening of the bottleneck and, hence, the weakening of mutual 
hindrance of the ribosomes. This trend of variation of flux with 
$\beta$ is also consistent with the gradual lowering of the average 
density profile with increasing $\beta$ as shown in the inset of 
Fig.~\ref{fig-fluxobc}(b). 

\section{\label{comparison}Comparison with vehicular traffic}

Our toy model is a simplified version of a more realistic model 
\cite{basuchowpre} which takes into account most of the important 
steps in the biochemical cycle of a ribosome during the elongation 
stage of protein synthesis. Another version, which is 
much simpler than even our toy model, has been studied extensively 
over the last four decades.
\cite{macdonald68,macdonald69,lakatos03,shaw03,shaw04a,shaw04b,chou03,chou04,dong}
In these earlier models, each ribosome is represented by a hard 
rod of length $\ell$ and the effects of the entire mechano-chemical 
cycle of a ribosome are captured by a single parameter $q$, which is 
the probability of hopping of the ribosome from one codon to the next 
per unit time. The trend of variation of $J$ with $\rho$ in those 
earlier models is qualitatively similar to that observed in 
our toy model (see Fig.~\ref{fig-fdpbc}). In the special case 
$\ell = 1$ the hard rods reduce to particles of unit size and 
the earlier models of ribosomal traffic become equivalent to the 
totally asymmetric simple exclusion process (TASEP) \cite{derrida}. 
In fact, TASEP is the simplest model of systems of interacting 
self-propelled particles.  \cite{sz,schuetz} 

It is known that for periodic boundary conditions, the exact 
expression for the flux $J$ in the TASEP is given by\cite{sz,schuetz}
\begin{equation}
J = q \rho (1-\rho). 
\label{eq-asep}
\end{equation}
For our toy model in the special case for which $\ell =1$ 
and $k_{\rm eff} \rightarrow \infty$, but $\omega_{fs} = q$ is nonzero 
and finite, $\Omega_{fs} \rightarrow 0$ and, consequently, the 
approximate expression (\ref{3:7}) for the flux reduces to Eq.~(\ref{eq-asep}). 

TASEP and its various extensions have been used successfully over 
the last two decades to model various aspects of vehicular traffic 
\cite{css,cnss} as well as many traffic-like phenomena in biological 
systems.\cite{polrev,tgf06,physica,lipo3,frey3,santen,popkov1}
Our toy model can be viewed also as a biologically motivated extension 
of TASEP to an exclusion process for extended particles with 
``internal states.''\cite{reichenbach,tabatabei}

A statistical distribution which is used widely to characterize the 
nature of vehicular traffic is the distance-headway distribution. 
In vehicular traffic, the distance-headway is defined by the spatial gap between 
two successive vehicles.\cite{css} In any particle-hopping
model, the number of empty sites in front of a vehicle is taken to be 
a measure of the corresponding distance-headway.\cite{chowstin,ss} For ribosome 
traffic we define the distance-headway as the number of the codons in between 
two successive ribosomes that are not covered by any ribosome. In 
the steady state of our toy model the distance-headway distribution is expected 
to be independent of the detailed internal biochemical dynamics. 
Therefore, the distance-headway distribution in our toy model is identical to 
that derived earlier for a TASEP-like model for 
ribosome traffic.\cite{shaw03} The expression is particularly simple in the special 
case $\ell = 1$. We leave it as a exercise (see Problem~3)
because it can be written down directly on purely physical grounds. 

\section{\label{conclusion}Summary and conclusions}
We have presented a simplified version of protein synthesis by ribosomes 
and analyzed our toy model using some elementary methods of statistical physics 
that are accessible to undergraduate students. This model captures the essential steps in the 
mechano-chemical cycle of each individual ribosome as well as the 
steric interactions between ribosomes that move simultaneously along 
the same mRNA template strand. In particular, we have reported the 
rates of protein synthesis and the average density profiles of 
ribosomes on their mRNA templates. 

We have investigated how the rate of protein synthesis is affected 
by the availability of the cognate tRNA molecules. We have demonstrated 
that, with the increase of the corresponding rate constant $\omega_{a}$, 
the flux saturates when the availability of cognate tRNA is no longer 
the rate limiting step in the synthesis of proteins.

The collective movement of ribosomes during protein synthesis is 
sometimes referred to as ribosome traffic because of its close 
superficial similarities with vehicular traffic on highways. We 
have discussed these similarities and crucial differences to 
put our work in a broader perspective.

For simplicity we have ignored 
the effects of sequence inhomogeneities of real mRNA tracks on 
which ribosomes move. It is straightforward to extend our model to 
take into account the actual sequence of codons on a given mRNA.
The simplest way\cite{basuchowpre} to capture the sequence 
inhomogeneity is to assume that the rate constant $\omega_{a}$ 
is site-dependent, that is, dependent on the codon species. By using 
this assumption, we have computed the rate of protein 
synthesis when two specific genes of a particular strain of 
{\it Escherichia~coli} bacteria are expressed. The lower flux 
observed for real genes, as compared to that for a homogeneous mRNA, 
is caused by the codon specificity of the available tRNA molecules.

The dynamics of interacting ribosomes during protein synthesis may be viewed as a biologically motivated extension 
of TASEP.\cite{sz,schuetz} These systems are never in thermodynamic 
equilibrium, but can attain non-equilibrium steady-states. The 
physical properties of models of interacting self-propelled 
particles have been investigated extensively in the recent years 
using concepts and techniques of non-equilibrium statistical mechanics 
\cite{sz,schuetz}.

\appendix

\section{Algorithm for the simulation of the toy model with periodic boundary conditions}
\noindent {\it Step 1 (Initialization)}

\begin{enumerate}[(a)]

\item Label the lattice sites by the integers $1,2,\ldots,L$ from left 
to right and 
assign occupation number $0$ to each site.

\item Put $N$ ribosomes, each of length $\ell$, randomly on the lattice without overlap.

\item Change the occupation number of the lattice sites covered by the left edge
of each ribosome to $1$.

\item To each ribosome, assign the chemical state $\mu = 1$.
(Alternatively, draw $\mu$ randomly from the three allowed integers 1, 2, 3.)

\end{enumerate}

\smallskip\noindent {\it Step 2 (Random selection)}. Using a random number 
generator, choose one of the $L$ sites and
if the corresponding occupation number is $1$, go to step 3; 
else, go to step 4.

\medskip \noindent {\it Step 3 (Updating mechano-chemical states)}. The chemical
state of the randomly selected ribosome is updated with the transition 
probability 
$W(1 \rightarrow 2) = 1 - e^{-\omega_{a} \Delta t}$, or 
$W(2 \rightarrow 3) = 1 - e^{-\omega_{f \ell} \Delta t}$, or 
$W(3 \rightarrow 1) = 1 - e^{-\omega_{fs} \Delta t}$, 
depending on whether it is in the state $\mu = 1$, $2$, or $3$, 
respectively. 
In the last case (i.e., corresponding to the transition $3 \rightarrow 1$), 
reset the occupation number of the old position of the ribosome 
to $0$ and that of its new position to $1$ if the transition takes place.

\medskip \noindent {\it Step 4}. Go to step 2. 
$L$ iterations of step 4 corresponds to one time step, 
which is equivalent to the real time $\Delta t$).

Iterate steps 2--4 up to ITERMX number of time steps 
so that the duration of the simulation in real time is 
${\tt ITERMX} \ast \Delta t$, where $\Delta t$ is a sufficiently small time interval.
The first {\tt ITST} time steps are used to ensure that the system settles to a steady-state.
The steady-state properties are computed over the next ${\tt ITERMX}-{\tt ITST}$ time steps.

\section{Suggested Problems}

\noindent Problem 1. (a) Derive analytically the expression 
for the flux $J$ in the steady-state 
for $N$ identical ribosomes of arbitrary size $\ell$ with periodic boundary conditions. Verify that 
the result reduces to Eq.~(\ref{3:7}) for $\ell = 1$. Plot $J$ against the coverage density 
\begin{equation}
\rho_{\rm cov} = N \ell/L = \rho \ell,
\end{equation}
and suggest a physical interpretation of the variation of $J$ 
with $\ell$.
(b) Imagine that the reverse transition from the state $2$ to 
the state $1$ is possible, that is, an aa-tRNA selected by the ribosome 
can detach prematurely from site A. Assume that the corresponding 
rate constant is $\omega_{p}$ and repeat the calculations of part (a) 
and interpret the results physically.

\medskip\noindent Problem 2. Write a computer program to simulate the model. 
Use open boundary conditions and compute 
$J$ and the average density profiles for arbitrary $\ell$.

\medskip \noindent Problem 3. Use purely heuristic arguments (without detailed 
calculations) to derive Eq.~(\ref{eq-asep}) for $J$ and 
the distance-headway distribution 
in the steady-state of TASEP for periodic boundary conditions. 

\begin{acknowledgements} 
We thank the editor and associate editor of AJP for very useful suggestions 
and advice, while editing this manuscript, which helped in improving the 
presentation of our work. 
This work has been supported (through DC) by the Council of Scientific and 
Industrial Research (CSIR), government of India. 
\end{acknowledgements}

%\newpage\section*{Figure Captions}
\newpage

\end{document}